\RequirePackage{fix-cm}
\documentclass[smallextended]{svjour3}       
\smartqed  
\usepackage{graphicx}
\begin{document}

\title{Tait equation revisited from the entropic and fluctuational points of view}



\author{E.B. Postnikov \and
        A.L. Goncharov \and \\
				V.V. Melent'ev}


\institute{E.B. Postnikov \at
              Department of Theoretical Physics, Kursk State University, Radishcheva st., 33, 305000 Kursk, Russia \\
              Tel.: +7-4712-56-14-39\\
              Fax: +7-4712-56-14-39\\
              \email{postnicov@gmail.com}           
           \and
           A.L. Goncharov \at
              Department of Theoretical Physics, Kursk State University, Radishcheva st., 33, 305000 Kursk, Russia
							\and
							V.V. Melent'ev \at
							Laboratory of Molecular Acoustics, Kursk State University, Radishcheva st., 33, 305000 Kursk, Russia 
}

\date{Received: date / Accepted: date}

\maketitle

\begin{abstract}
We consider the possibilities for prediction of liquids' density under pressure basing on the inverse reduced fluctuations parameter, which is directly connected with the isothermal compressibility. This quantity can be determined basing on the thermodynamical properties of  saturated liquid and it consists of only two constant parameters within a relatively wide region close to the melting points. It is confirmed by the comparison with the experimental data on n-alkanes that the derived expression is a quite reasonable estimator without a necessity to fit data along some parts of isotherms for different temperatures. At the same time the obtained formula: i) can be reduced to the form of the Tait equation and ii) the resulting Tait's parameters in this representation have a clear physical meaning as functions of the excess entropy, which determines the mentioned reduced fluctuations.

\keywords{Excess entropy \and Tait equation \and Volume fluctuations}
\end{abstract}

\section{Introduction}

The Tait equation
\begin{equation}
\frac{\rho-\rho_0}{\rho}=C\log\left[\frac{(P-P_0)}{B+P_0}+1\right]
\label{tait}
\end{equation}
is an effective approximation of the density along an isotherm for various classes of liquids 
(from simple to aggregate-forming ones), which is wide used during more then century; see the very comprehensive historical review \cite{Dymond1988tait}. 

At the same time, although the effectiveness of Eq.~(\ref{tait}) is out of question, the strict physical origin of the equations as well as the meaning of its coefficients is still a discussed question. Several attempts to clarify this problem are known: from earlier thermodynamic interpretations \cite{Ginell1961,Nanda1964,Jain1989} to more recent statistical physics approaches \cite{Bulavin1997,Fakhretdinov2004}. 

However, the cited works basically deal with the  approximations (mainly polynomial) of an equation of state for liquids. As a result, Tait' coefficients are explained through other adjusted parameters. As a result, their determination for a practical usage is still the problem of fitting known experimental data on the pressure and the density for a certain region and each temperature separately. Then, this fitted values can be successfully used for the extrapolation to much more larger pressures, or with the goal of analytical fit of given obtained data \cite{Dymond1988,Assael1994,Randzio1994,Ramos2006}. 

Recently, another approach has been proposed \cite{Goncharov2013} for the study of the coexistence curve of simple liquids: the analysis of the inverse relative fluctuations along this line. It has an advantage of the clear interconnection of this quantity with the structural (the excess entropy, the structure factor $S(q)$ at the wave number $q = 0$) and transport (self-diffusion) characteristics of a liquid as well as it can be supported by both direct acoustic and thermophysical measurements and lattice fluid simulations. 

Thus, the main goal of this work is to expand this approach for the case of single-phase liquids under pressure and to show that it is directly connected with the Tait equation providing therefore the structural physical arguments about the origin of Tait's coefficients and possible methods of their calculations from the first principles.

\section{Relative volume fluctuations and Tait parameters}

\subsection{Single phase under pressure: fluctuational point of view}

It has been argued in the recent work \cite{Goncharov2013} that the combination
\begin{equation}
\nu=\frac{\mu c^2}{\gamma RT},
\label{nu}
\end{equation}
where $\gamma$ and $c$ are the heat capacity ratio and speed of sound correspondingly, provides an important information on the structure of liquids and the strength of intermolecular interactions in a given medium along the coexistence curve. 
	
This parameter $\nu$ is the inverse ratio of relative volume fluctuation to its value in the hypothetical case where the substance acts an ideal gas for the same temperature-volume parameters and can be expressed through the isothermal compressibility as follows:
\begin{equation}
\frac{1}{\beta_T}=\nu \frac{RT}{\mu}\rho=\left(\frac{\partial \rho}{\partial P}\right)_T\rho.
\label{betanu}
\end{equation}

Therefore, we can consider the corollary of this relation for the prediction of single-phase density along an isotherm. Basing on the fact of independence of  volume and temperature fluctuations \cite{Landau}, the simple integration of Eq.~(\ref{betanu}) gives
\begin{equation}
P-P_0=\frac{RT}{\mu}\int\limits_{\rho_0}^{\rho}\nu(\rho)d\rho,
\label{pint}
\end{equation}
where $(\rho_0, P_0)$ is the referent state, say, a point on the coexistence curve or the density and pressure of a liquid under the atmospheric pressure. 

In the work \cite{Goncharov2013} we have determined that inverse relative fluctuations in simple real liquids (liquefied noble gases and coexisting vapours) and model lattice systems  have a strong exponential character within the relatively wide region close to the melting point that supports the ideas of free-volume theory and estimations by the methods of statistical physics related to the excess entropy. 

Now, we evaluate the same fitting procedure for the series of liquid n-alkanes using thermodynamical data from {\tt  NIST Chemistry WebBook} \cite{NIST}. Results of the approximation in the form $\nu(\rho)=\exp(k\rho-b)$ for the liquid branch is presented in  Table.~\ref{tablefit}. Substituting this exponential approximation into (\ref{pint}) and integrating it, we obtain the final expression for the prediction of liquid's density along an isotherm as a function of the pressure:
\begin{equation}
\rho=\rho_0+\frac{1}{\kappa}\log\left[\frac{\kappa\mu}{\nu(\rho_0)RT}(P-P_0)+1\right].
\label{rhopredict}
\end{equation}

\begin{table}
\label{tablefit}
\caption{Parameters of approximation of the inverse reduced fluctuations within the interval of validity of an exponential fit (they are presented for the temperature and density ranges). The density in the last column is measured in $kg\cdot m^{-3}$.}
\begin{tabular}{c|c|c|c}
\hline
Substance&$T\, (K)$&$\rho\, (kg\cdot m^{-3})$&$\log(\nu)$\\
\hline
Hexane&189.38--332.93&751.35--621.82&$0.0124\rho-4.715$\\
Heptane&195.07--350.61&765.42--633.68&$0.0125\rho-4.715$\\
Octane&228.72--382.26&753.76--627.34&$0.0124\rho-4.730$\\
Nonane&232.82--395.88&765.63--634.02&$0.0123\rho-4.650$\\
Dodecane&277.41--449.02&761.09--629.71&$0.0124\rho-4.687$\\
\hline
\end{tabular}
\end{table}

To test a quality of approximation, we compare the calculation results with the known experimental data on the density of n-alkanes under the high pressure.

\begin{figure}
\includegraphics[width=\textwidth]{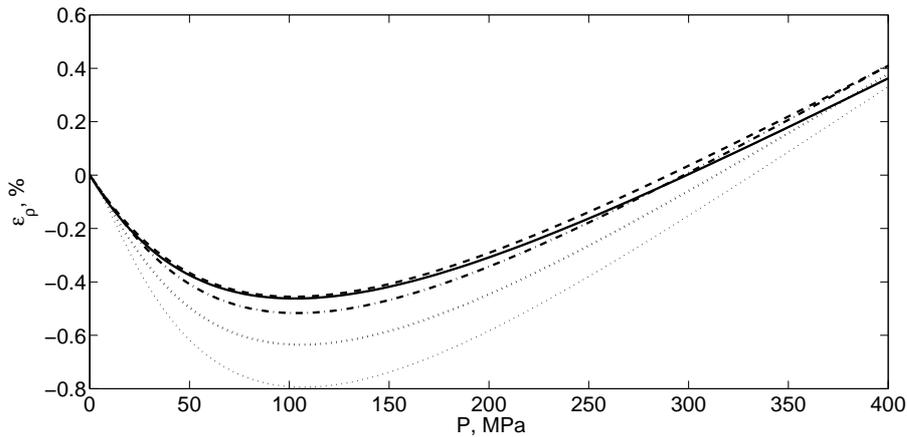}
\caption{The relative errors between the values of density predicted using Eq.~(\ref{rhopredict}) and the fitted experimental data \cite{Randzio1994} for the isotherms: $T=283.15\, K$ (solid line), $T=303.15\,K$ (dashed line), $T=323.15\,K$ (dash-dotted line), $T=343.15\,K$ (thick dotted line), $T=363.15\,K$ (thin dotted line).}
\label{fighexan}
\end{figure}

First of all, we use the data on n-hexane, which are  presented and scrutiny analyzed in the work \cite{Randzio1994}. Note that the authors use namely the Tait equation is an approximant (certainly, with the specially adjusted $C$ and $B$ for each isotherm). This fact allows us to draw the relative error of approximation between these fitted experimental data (with the claimed standard deviation $0.12\%$) and the data predicted by Eq.~(\ref{rhopredict}) with the parameters $\kappa$, $b$ listed in Table.~\ref{tablefit} as a continuous curve. The results for the series of isotherms are plotted in Fig.~\ref{fighexan}. 

One can see that the relative error of approximation does not exceed $0.5\%$ for the temperature region, for which the exponential fit of the inverse reduced fluctuations was evaluted (see Table.~\ref{tablefit}), and these curves are quite close to each other. Moreover, even for two isotherms with the temperatures larger than $332.93\,K$ (dotted lines in Fig.~\ref{fighexan}) the error is limited to $1\%$ range up to 400 MPa.

For several other n-alkanes, the comparison is carried out using the raw experimental date from the the article \cite{Boelhouwer1960}. The results are presented in Table.~\ref{alctest}.

One can see that the quality of prediction is still reasonable (with the accuracy up to 5\%) for all substances and the gauge pressure interval up to 100 MPa (and having in mind the accuracy of data from \cite{Boelhouwer1960}, which may be lower than detailly analysed one for n-hexane in \cite{Randzio1994}).

Thus, the simple formula~(\ref{rhopredict}) with the parameters depending on the saturated values only, gives a worthy prediction for the density under pressure  even if it cannot serve as high-accurate interpolation of the actually given data. Additionally, it has an advantage of the clear physical meaning of its coefficients, as it will be discussed below.

\begin{table}
\label{alctest}
\caption{The comparison of predicted by Eq.~(\ref{rhopredict}) and experimental \cite{Boelhouwer1960} data for liquid n-alkanes under pressure.}

\includegraphics[height= \textheight]{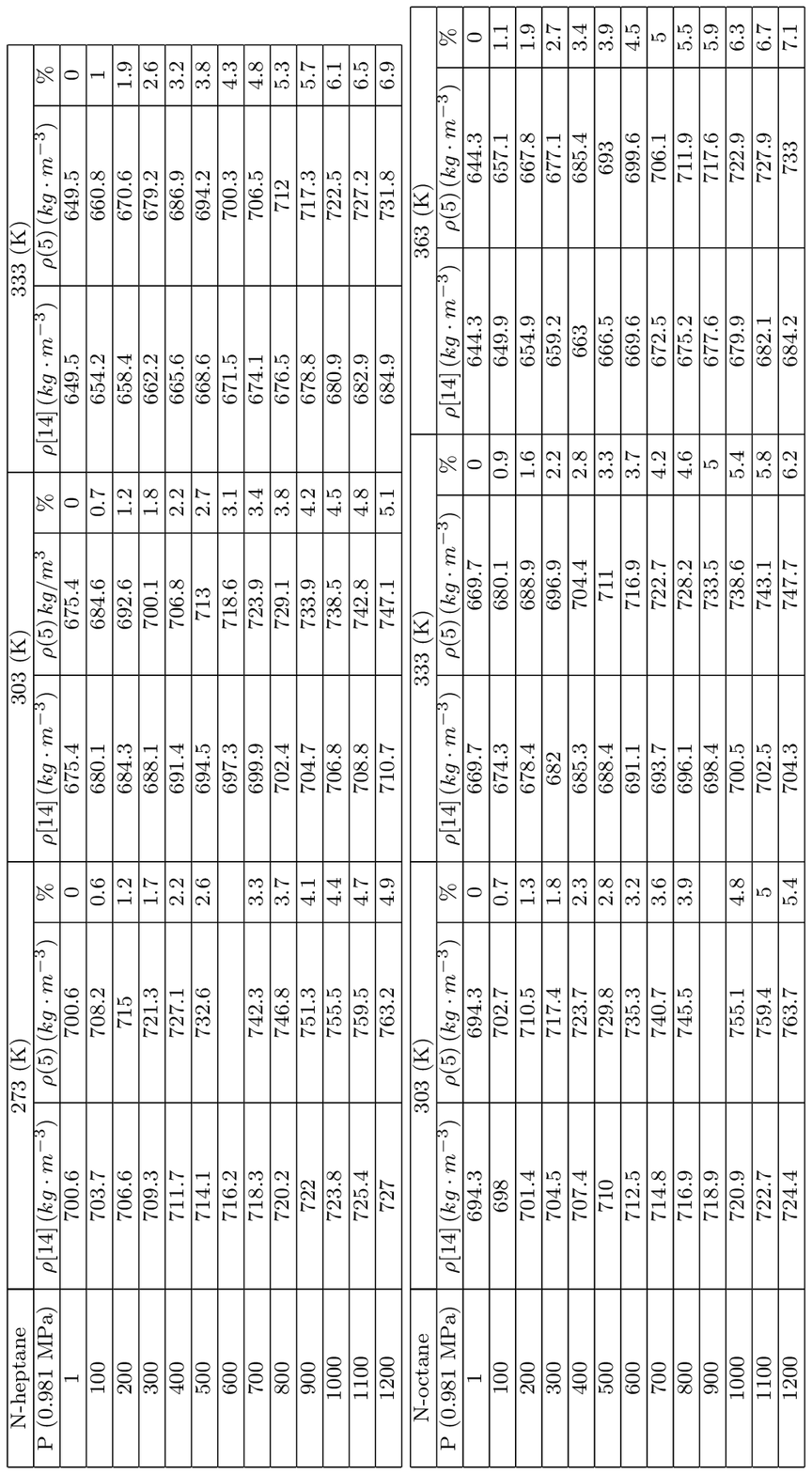}

\end{table}

\includegraphics[height= \textheight]{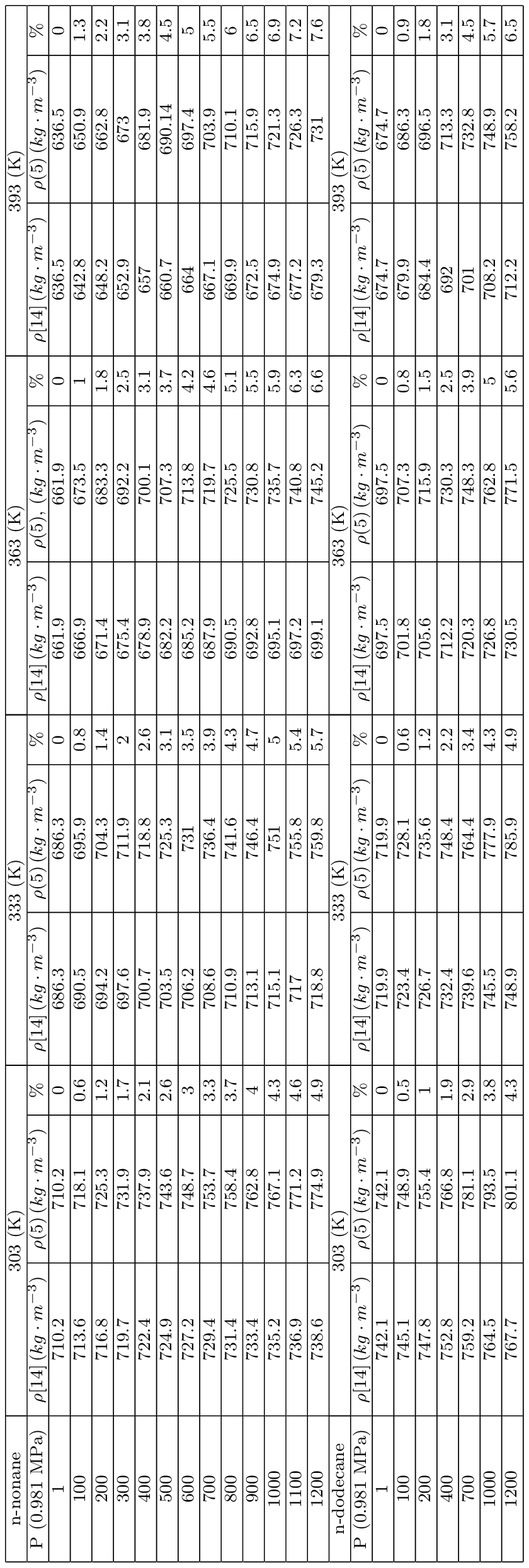}

\subsection{Fluctuational predictor and the Tait equation}

Let us express the factor $\kappa^{-1}$ before the logarithm in Eq.~(\ref{rhopredict}) from the exponential expression for the inverse relative fluctuations. Whereupon it will consist of the reference density as a general multiplier, the equation (\ref{rhopredict}) can be transformed to the expression for the relative density change,
\begin{equation}
\frac{\rho-\rho_0}{\rho}=\frac{1}{\log(\nu(\rho_0))+b}
\log\left[\frac{(P-P_0)}{\left\{\frac{\nu(\rho_0)RT}{\kappa\mu}-P_0\right\}+P_0}+1\right],
\label{flucttait}
\end{equation}
which formally coincides with the Tait equation (\ref{tait}).

Thus, the Tait coefficients actually depend on the position of the referent point  on the liquid-vapour coexistence curve and are the functions of the saturated density (or the temperature since they are functionally connected):
\begin{equation}
C=\frac{1}{\log(\nu(\rho_0))+b},\quad B=\frac{\nu(\rho_0)RT}{\kappa\mu}-P_0.
\label{CB}
\end{equation}

These expressions allow to reveal the structural origin of this behaviour of coefficients. Previously it has been shown \cite{Goncharov2013} that the parameter of inverse reduced volume fluctuations $\nu$ is equal to the ratio of the self-diffusion coefficient of a dense medium having the properties of an ideal gas to an actual self-diffusion coefficient, the parameter, which satisfied Dzhugutov's universality \cite{Dzugutov1996}:
$$
\nu(\rho_0)^{-1}=D^*(\rho_0)=A\exp\left[S_2(\rho_0)\right],\quad A=\mathrm{const},
$$ 
where $S_2$ is the two-body approximation (i.e. using the radial distribution function $g(r)$ only)
\begin{equation}
S_2(\rho_0)=-2\pi\rho_0\int_0^{\infty}\{g(r)\log[g(r)]-[g(r)-1]\}r^2dr
\label{entropy}
\end{equation}
 of the excess entropy $S_{ex}\equiv S-S_{ig}$, the internal configurational quantity  
defined \cite{Rosenfeld1977} as the difference between the total entropy $S$ of a fluid, and the entropy $S_{ig}$ of an ideal gas under the same PVT conditions. 

The integral in Eq. (\ref{entropy}) for practically uniform dense media (beyond the first coordination shell)  in absence of sharp structural transition is almost constant (see \cite{Yan2008} and more detailly \cite{Goncharov2013}). Thus,  the excess entropy is a linear function of density for liquids in the state close to the melting point. Our constant $b$ can expressed via Dzhugutov's one (which is connected with the Enskog collision rate) as $b=-\log(A)$.

As a result, the Tait coefficients read
$$
C(\rho_0)=-S_2(\rho_0)\quad \mathrm{and}\quad  
B(\rho_0)= -\frac{\rho_0RT}{\mu}\frac{1}{AS_2e^{S_2(\rho_0)}}-P_0
$$
as the functions of the excess entropy of a system, which is negative for dense fluids. 

However, although the expressions above connect parameters with the inner structural characteristics, their practical estimation is more reasonable using the direct adjustment to the experimental data along the coexistence curve as it has been done above.

\section{Conclusion}

Thus, we show in this work that the two parameters determining the behaviour of the inverse reduced volume fluctuations are enough to characterize the density of liquids under high pressure in a single-phase state for a wide interval of parameters of state. The microscopic   interpretation of this fact can be the following: the structure of liquid within this range does not have any structure transitions and can be represented as a simple uniform squeezing. 
The direct interconnection of the mentioned parameters with the excess entropy confirms this interpretations.

As well, the proposed formula for density prediction under pressure along an isotherm provides the satisfactory accuracy. It can be transformed into the form of the Tait equation but its practical applicability does not require a fitting of parameters within the single-phase region for various temperatures separately.  Therefore, the proposed representation provides the prognostic opportunities for a larger set of substances due to much more wider availability of data on saturated thermodynamical values.

Finally, the revealed interconnection between Tait's parameters and the excess entropy opens the opportunities for their calculation from the first principles since the last quantity can be expressed through the radial distribution function, the methods of calculation of which are developed, see e.g. \cite{Touba1997}. 

\begin{acknowledgements}
We are grateful to  Prof. Yu.A. Neruchev (Research Center for Condensed Matter Physics, Kursk State University) and to the participants of 10th Winter Workshop on Molecular Acoustics, Relaxation and Calorimetric Methods (04-07.03.2014, Szczyrk, Poland), where this work has been presented, for fruitful discussions. The work is supported by the grant No.~1391 of the Ministry of Education and Science of the Russian Federation within the basic part of research funding No. 2014/349 assigned to Kursk State University.
\end{acknowledgements}


\begin{thebibliography}{10}

\bibitem{Dymond1988tait}
J.H. Dymond, R.~Malhotra, Int. J. Thermophys. {\bf 9}, 941 (1988)

\bibitem{Ginell1961}
R.~Ginell, J. Chem. Phys. \textbf{34}, 1249 (1961)

\bibitem{Nanda1964}
V.S. Nanda, R.~Simha, J. Chem. Phys. \textbf{41}, 1884 (1964)

\bibitem{Jain1989}
R.K. Jain, R.~Simha, Macromolecules \textbf{22}, 464 (1989)

\bibitem{Bulavin1997}
L.A. Bulavin, V.M. Sysoev, I.A. Fakhretdinov, Theor. Math. Phys. \textbf{111},
  771 (1997)

\bibitem{Fakhretdinov2004}
I.A. Fakhretdinov, E.R. Zhdanov, High Temperature \textbf{42}, 396 (2004)

\bibitem{Dymond1988}
J.H. Dymond, R.~Malhotra, J.D. Isdale, N.F. Glen, J. Chem. Thermodynamics
  \textbf{20}, 603 (1988)

\bibitem{Assael1994}
M.J. Assael, J.H. Dymond, D.~Exadaktilou, Int. J. Thermophys. \textbf{15}, 155
  (1994)

\bibitem{Randzio1994}
S.L. Randzio, J.P.E. Grolier, J.R. Quint, D.J. Eatough, E.A. Lewis, L.D.
  Hansen, Int. J. Thermophys. \textbf{15}, 415 (1994)

\bibitem{Ramos2006}
M.~Ramos-Estrada, G.A. Iglesias-Silva, K.R. Hall, J. Chem. Thermodyn.
  \textbf{38}, 337 (2006)

\bibitem{Goncharov2013}
A.L. Goncharov, V.V. Melent'ev, E.B. Postnikov, Eur. Phys. J. B \textbf{86},
  357 (2013)

\bibitem{Landau}
L.D. Landau, E.M. Lifshitz, \emph{Statistical Physics. Part 1.} (Oxford:
  Pergamon Press, 1980)

\bibitem{NIST}
The NIST Chemistry WebBook provides access to data compiled and distributed by
  NIST under the Standard Reference Data Program, http://webbook.nist.gov

\bibitem{Boelhouwer1960}
J.W.M. Boelhouwer, Physica \textbf{26}, 1021 (1960)

\bibitem{Dzugutov1996}
M.~Dzhugutov, Nature \textbf{381}, 137 (1996)

\bibitem{Rosenfeld1977}
Y. Rosenfeld, Phys. Rev. A \textbf{15}, 2545 (1977).

\bibitem{Yan2008}
Z.~Yan, S.V. Buldyrev, H.E. Stanley, Phys. Rev. E \textbf{78}, 051201 (2008)

\bibitem{Touba1997}
H.~Touba, G.A. Mansoori, Int. J. Thermophys. \textbf{18}, 1217 (1997)

\end{thebibliography}

\end{document}